\documentclass[11pt]{article}
\usepackage{graphicx}
\usepackage{epsfig}
\usepackage{axodraw}
\usepackage{color}

\usepackage{amssymb,amsmath}
\usepackage{graphicx}  
\newcommand{\beq}{\begin{equation}}
\newcommand{\eeq}{\end{equation}}
\newcommand{\be}{\begin{equation}}
\newcommand{\ee}{\end{equation}}
\newcommand{\bea}{\begin{eqnarray}}
\newcommand{\eea}{\end{eqnarray}}

\fontencoding{T1}
\fontfamily{garamond}
\fontseries{m}
\fontshape{it}
\fontsize{13}{15}
\selectfont

\begin{document}

\title{The Charged $Z(4433)$: Towards a New Spectroscopy}
\author{L.~Maiani$^a$, A.D. Polosa$^b$, V.~Riquer$^b$\\
$^a$Dip. Fisica, Universit\`a  di Roma ``La Sapienza'' and INFN, Roma, Italy\\
$^b$INFN, Sezione di Roma, Roma, Italy}
\maketitle
\begin{abstract}
We identify the newly found $Z(4433)$ with the first radial excitation of the  tetraquark basic supermultiplet to which $X(3872)$ and $X(3876)$ belong.
Experimental predictions following from this hypothesis are spelled out.
\newline
{\bf Keywords} Hadron Spectroscopy, Exotic Mesons, Heavy Meson Decays \newline
{\bf PACS} 12.39.Mk, 12.40.-y, 13.25.Jx
\end{abstract}

The Belle Collaboration has reported the observation of a new resonance-like state, $Z(4433)$, via the decay~\cite{belle}: 
\be
Z(4433)\to\psi(2S)+\pi^\pm
\label{zdecay}
\ee

Given the  narrow width  ($\Gamma\sim 44$~MeV), the  natural interpretation of $Z(4433)$ is that of diquark-antidiquark state with flavors $[cu][\bar c\bar d]$.
In this case, the decay~(\ref{zdecay}) indicates  negative charge conjugation for the neutral partner and, assuming S-wave decay, $J^{PC}=1^{+-}$.

We have proposed in~\cite{mppr} to identify the $X$ states discovered by Belle and BaBar~\cite{X1} with the lowest lying tetraquarks. Recently, the observation 
that there are in fact two, close in mass, neutral states
$X(3872)$ and $X(3876)$~\cite{X2},  has led us to propose~\cite{noi}:
\bea
&&X(3872)=X^0_{u\bar u}(1^{++};1S)\notag\\
&&X(3876)=X^0_{d\bar d}(1^{++};1S)
\eea

The tetraquark hypothesis implies a proliferation of $X$ states, already indicated by the anomalous $p_\perp$ distribution of $\psi$ in $B$ decay~\cite{bigi}, that experiments are starting to unveil~\cite{faccini}.

In~\cite{mppr}, we have studied the lowest lying tetraquark supermultiplet 
within the constituent quark model. Besides those associated with $X(3872)/X(3876)$, the supermultiplet contains two states with quantum numbers $J^{PC}=1^{+-}$. 
Properties of the highest $1^{+-}$ state were estimated in~\cite{mppr} to be: 
\bea
&&X^{+}_{u\bar d}(1^{+-};1S):\notag\\
&&M\sim 3880; \;\;\;  X^{+}_{u\bar d}(1^{+-};1S)\to \psi(1S)+\pi^+ ~{\rm or}~\eta_c(1S)+\rho^+
\label{unopm}
\eea

It is interesting to observe that the mass difference between $Z(4433)$ and the expected 
$X^{+}_{u\bar d}(1^{+-}; 1S)$ is close to the mass difference $M_{\psi(2S)}-M_{\psi(1S)}\sim 590$~MeV so as to suggest that, in fact, Belle may have observed the first radial excitation of the $1^{+-}$ tetraquark. The decay in $\psi(2S)$ rather than $\psi(1S)$ lends credibility to the statement.

In this note, we elaborate on a few testable predictions that would follow from the assignment:
\be
Z(4433)=X^{+}_{u\bar d}(1^{+-}; 2S).
\label{crux}
\ee

\paragraph{The $1^{+-}, 1S$ charged state.} A crucial consequence of~(\ref{crux}) is that a charged state decaying in $\psi+\pi^\pm$ or $\eta_c+\rho^\pm$ should be found around $3880$~MeV, according to~(\ref{unopm}).

\paragraph{Neutral partners of $Z(4433)$.} It  goes almost without saying that $Z(4433)$ must have neutral partners according to the scheme: 
\bea
&&X^0_{u\bar u, d\bar d}(1^{+-};2S):\notag\\
&&M\sim M_{Z(4433)}\pm {\rm few~MeV};\;\;\; X^0_{u\bar u, d\bar d}(1^{+-};2S)\to \psi(2S)+\pi^0/\eta ~{\rm or}~\eta_c(2S)+\rho^0/\omega
\eea

\paragraph{The $2S$ supermultiplet.} 

There are indications of a $X$ state with decay in $D^*\bar D^*$ produced in:
\be
e^+e^-\to \psi~ D^*\bar D^*
\ee
at a mass $M=4160$~MeV.
The charge conjugation of such a state is $+1$ so that $J^{PC}=0^{++}~{\rm  or}~ 2^{++}$ if S-wave decay is assumed.
This particle could be identified with $X_{q\bar q^\prime}(0^{++};2S)$ although a conventional classification as $\eta_c(3S)$ is also viable at present.

One should also find the $1^{++}$ states:
\bea
&&X_{q\bar q^\prime}(1^{++};2S):\notag\\
&&M\sim 4200-4300;\;\;\; X_{q\bar q^\prime}(1^{++};2S)\to D^{(*)}D^{(*)}
\eea
while $2^{++}, 2S$ states should be around $4600$~MeV.

\paragraph{The baryon-antibaryon threshold.} The  tetraquark mesons have a strong affinity for the decay into baryon-antibaryon channels~\cite{bibbar}, \cite{mppr}. The thresholds  are:
\be
2M_{\Lambda_c^+}=4572~{\rm MeV};\;\;\;\;\; M_{\Lambda_c^+}+M_{\Sigma_c^0}=4739~{\rm MeV}
\ee
for hidden-charm, neutral and charged tetraquarks, respectively.
$X$ mesons with masses above this limit are expected to be considerably broader than the presently observed ones, which would provide another conclusive test  of the present scheme.
\paragraph{Acknowledgements}
We thank R. Faccini for interesting discussions and useful information.
Part of this work was performed during the International School Ettore Majorana, Erice, Sicily.


\begin{thebibliography}{99}

\bibitem{belle}
    [Belle Collaboration],
  arXiv:0708.1790 [hep-ex].
\bibitem{mppr}
  L.~Maiani, F.~Piccinini, A.~D.~Polosa and V.~Riquer,
  Phys.\ Rev.\  D {\bf 71}, 014028 (2005)
  [arXiv:hep-ph/0412098].


\bibitem{X1} S.~K.~Choi {\it et al.}  [Belle Collaboration],
  Phys.\ Rev.\ Lett.\  {\bf 91}, 262001 (2003)
  [arXiv:hep-ex/0309032];
  B.~Aubert {\it et al.}  [BABAR Collaboration],
  Phys.\ Rev.\  D {\bf 73}, 011101 (2006)
  [arXiv:hep-ex/0507090].

\bibitem{X2} G.~Gokhroo {\it et al.},
  Phys.\ Rev.\ Lett.\  {\bf 97}, 162002 (2006)
  [arXiv:hep-ex/0606055]; P. Grenier (for the BELLE and BABAR Collaborations), {\it Charm and charmonium spectroscopy at B-factories}, Moriond QCD 2007, March 17th-24th.

\bibitem{noi}
  L.~Maiani, A.~D.~Polosa and V.~Riquer,
  arXiv:0707.3354 [hep-ph].

\bibitem{bigi}
  I.~Bigi, L.~Maiani, F.~Piccinini, A.~D.~Polosa and V.~Riquer,
  Phys.\ Rev.\  D {\bf 72}, 114016 (2005)
 [arXiv:hep-ph/0510307].

\bibitem{faccini}
See e.g. R.~Faccini in Lepton Photon 2007.

\bibitem{bibbar}
  L.~Maiani, F.~Piccinini, A.~D.~Polosa and V.~Riquer,
  Phys.\ Rev.\ Lett.\  {\bf 93}, 212002 (2004)
  [arXiv:hep-ph/0407017]; 
  G.~C.~Rossi and G.~Veneziano,
  Phys.\ Lett.\  B {\bf 597}, 338 (2004)
  [arXiv:hep-ph/0404262] and references therein.





\end{thebibliography}
\end{document}